\documentstyle[twocolumn,twoside]{article}
\input epsf.tex
\newcommand{\h}{\linebreak \hspace*{3ex}}
\newcommand{\hb}{\\ \hspace*{2ex}}

\begin{document}
\title{EVOLUTION OF A CORRELATION FUNCTION OF QSOs AND
THE INITIAL POWER SPECTRA OF DENSITY FLUCTUATIONS}
\author{Yu.B.\,Chornij, B.S.\,Novosyadlyj\\[2mm]
\begin{tabular}{l}
\centerline{Astronomical Observatory, L'viv State University}\hb 
\centerline{Kyryla and Mephodia str.8, L'viv 290005 Ukraine}\hb 
\centerline{{\em yuch@astro.franko.lviv.ua, novos@astro.franko.lviv.ua}}\\
\end{tabular}
}
\date{}
\maketitle

ABSTRACT.
The evolution of two point space correlation function of QSOs ($\xi_{QQ}(r,z)$) is analyzed
in the framework of the theory of the large scale structure formation.
For given cosmological models the agreement between
theoretical predictions and observational data on $\xi_{QQ}(r,z)$
is achieved by determination of scales of structures $ R (z) $, in
which could be formed quasars at different redshifts $ z $.
It is shown that quasars could be formed in the nucleus of massive galaxies or groups of galaxies 
owing to merges at their central region. It depends on an amplitude 
of the power spectrum of the initial matter density 
fluctuations  at galaxy-cluster scales.\\[1mm]


{\bf 1. Introduction}\\[1mm]

In the problem  of the large-scale structure formation the QSOs present 
a special interest because they cover almost all epoch of evolution of 
the Universe after cosmological recombination. The most physically motivated
explanation of quasar phenomena is the accretion of gas and stars 
into the massive black holes in the center of galaxies. It would take place at the 
early stage of galaxy evolution or later as a result of collision or merging of
galaxies in the groups or clusters.
Therefore, the theoretical precalculation of spatial two-point 
correlation functions of QSOs ($ \xi _ {QQ} (r, z)) $ has an important 
meaning for testing  the cosmological scenarios of the formation of the large-scale 
 structure of the Universe. 

The analysis of quasar surveys 
(Iovino et al. 1991; Andreani, Cristiani 1992; Mo, Fang 1993;
Komberg, Kravtsov 1995; Fang 1998; La Franca et al. 1998; 
Cristiani et al. 1998 and references therein) indicates on the clustering  of QSOs 
higher than galaxies and lower than rich clusters.
The observable quasar correlation function  can be approximated by the expression
\mbox{$\xi_{QQ}(r,z)=(r/r_0^Q)^{-\gamma}(1+z)^{-\alpha}$} with $\gamma\approx 1.8$
and $r_0^{Q}\approx 6-10h^{-1}$Mpc.
As it was shown  by Iovino et al. 1991, 
Mo, Fang 1993 and  Komberg, Kravtsov 1995 the amplitude of 
$\xi_{QQ}(r,z)$ decreases with increasing of redshift $ z $ ($\alpha>0$). 
But they give the 
different rate (values of $\alpha$) of evolution. 
Andreani, Cristiani 1992 did not reveal an essential evolution of $\xi_{QQ}(r,z)$. 

Contrary, taking into account the last data on QSO space distribution La Franca et al. 1998 and 
Cristiani et al. 1998 show that amplitude of $ \xi_{QQ}(r,z)$ increases with $z$.
According to their results 
\begin{equation}
\label{xi_QQ_obs}
\bar \xi_{QQ}^{obs}(15h^{-1} Œ¯ª,z)=
(r_0^Q/r)^{1.8}(1+z)^{-(1.2+\epsilon)},
\end{equation}
where the "fitting-parameter" $ \epsilon = -2.5\pm1.0 $. Here $ \bar\xi_{QQ}(15h^{-1}Mpc,z)$
is integrated correlation function defined as 
\begin{equation}
\label{xi_QQ_obs_int}
\bar \xi_{QQ}(r,z)=
\frac{3}{r^3}\int\limits_0^{r}x^2\xi_{QQ}(x,z)dx.
\end{equation}

The goal of this paper is  theoretical interpretation of such evolution of
$ \bar\xi _ {QQ} (15h ^ {-1} Mpc, z) $ within the framework of cosmological models
with the given initial power spectra $P(k)$ and existing assumptions about 
physical nature of quasars.  \\ [2mm]

{\bf 2. Quasars correlation function}\\[1mm]

We assume that the QSOs are manifestation  of short-term active processes, 
which occur in nucleuses of massive galaxies at the early stage of their evolution
or later one owing the merging of galaxies 
in their groups or clusters (Haehnelt 1993; Haehnelt, Rees 1993; Loeb, Rasio 1994;
Kontorovich 1996; Efstathiou, Rees 1998).  
According to the most elaborated scenario of the large scale structure formation 
the galaxies, their groups and clusters are formed in the peaks of the random Gaussian field
of the matter density linear perturbations owing to gravitational instability.
So, the amplitude of relative density fluctuations  
\mbox {$\delta \equiv \frac {(\delta\rho)} {\rho} = \nu\cdot\sigma (R) $}
($\nu$ is the height of peaks, $ \sigma(R) $ is the root mean square of 
density fluctuation at corresponding scale $R$)
are normally distributed.
The quasar stage in an evolution of such structures begins 
at some moment of time $ t $ (or redshift $z$), when massive black hole at center
of galaxies is formed or close galaxies in most dense region of groups or clusters
are merged.  
We can neglect an interval of time $ \Delta t $, during which the quasar mechanism 
is formed, because it is essentially smaller than cosmological time $ t $ 
$ \left (\Delta t \ll t\right) $ as it was shown by Efstathiou, Rees 1988,
Haehnelt 1993, Haehnelt, Rees 1993, Loeb, Rasio 1994 and
Kontorovich 1996. It means that the moment of 
beginning of quasar stage is close to the moment of appearence of the singularity
in the Tolman solution for the collapse of dust-like cloud. In the initial random 
field of the density fluctuations the peaks of corresponding scale and amplitude 
result into such collapsing clouds.
 The values of $ \delta $ for  fluctuations which collaps at redshift z is equal to
  $ \delta (z) = \delta _ c \cdot (z + 1) $. 
In a case of spherical - symmetric fluctuations - $ \delta _ c = 1.69 $,  
\mbox {nonspherical -} $ \delta _ c = 1.33 $  
(Efstathiou, Rees 1988, Nusser, Silk 1993).

In the framework of the theory of the random  Gaussian fields (Bardeen et al. 1986) 
the two-point correlation function of peaks in which the QSOs will be formed is  following:
\begin{equation}
\label{xi_pp}
\xi_{QQ}(r,z)=\left(\sqrt{\xi_{QQ}^s(r,z)}\;+\sqrt{\xi_{\delta\delta}(r,z)}\right)^2,
\end{equation}
where $ r $ - distance between them in comoving coordinates, $ \xi_{QQ}^{s} (r, z) $ is the 
statistical component of correlation function of peaks in which 
the QSOs on given $ z $ are formed, $ \xi_{\delta\delta} (r, z) $ is the correlation function
of matter density fluctuations, it is the dynamical component. 

For model with given initial power spectrum  of density fluctuations $ P(k) $
the correlation function $ \xi_{\delta\delta} (r, z) $ is equal
\begin{equation}
\label{xi_p}
\xi_{\delta\delta}(r,z)=\frac{1}{2\pi^2}\cdot\int\limits_0^{\infty}\\
k^2\cdot\frac{P\left(k,R\right)}{(1+z)^2}\cdot\frac{sin(kr)}{kr}\,dk,
\end{equation}
where $P\left(k,R\right)=P(k)\cdot W^2(kR)$, $W(kR)=exp\left(-\frac{1}{2}\cdot k^2\cdot R^2\right)$
is the Gaussian window function for the scale $ R $, 
which corresponds to scale of fluctuations. The height of peaks of density fluctuations, 
in which the QSOs will be formed on fixed $ z $, is defined as:
\begin{equation}
\label{nu_z}
\nu(z)=\frac{\delta_c\cdot(1+z)}{\sigma(R)}\;.
\end{equation}
Then $\xi_{\delta\delta}(r, z)$ can be presented as follows:
\begin{equation}
\label{xi_p}
\xi_{\delta\delta}(r,z)
=\left(\frac{\delta_c}{\nu(z)}\right)^2\cdot x(r),
\end{equation}
where $x(r)=\xi_{\delta\delta}(r)/\xi_{\delta\delta}(0)$,
$\xi_{\delta\delta}(r)\equiv\xi_{\delta\delta}(r,z=0)$.

The expression for $ \xi_{QQ}^{s} (r, z) $ was obtained by  
Novosyadlyj, Chornij 1998 by the method like Kaiser 1984 one
for correlation function of 
rich clusters but taking into account that QSO phenomenon is
short-term stage after collapse (or merging) in center of corresponding
proto-object. It has the following view:
\begin{equation}
\label{xi_qq}
\xi_{QQ}^s(r,z)
=\frac{1}{\sqrt{1-x^2(r)}} \cdot exp\left(\frac{\nu^2(z)\cdot x(r)}{1+x(r)}\right)-1.
\end{equation}
Substituting expression (\ref{xi_p}),(\ref{xi_qq}) in expression (\ref{xi_pp}),
we have complete expression for correlation function of quasars $\xi_{QQ}(r, z) $ which we
analyze here.\\[2mm]

{\bf 3. Evolution of correlation function of quasars}\\[1mm]

At the beginning let's consider the case when the  QSOs are formed in structures 
of the same fixed scale $ R $, which does not depend on cosmological time $ t $.
For this we will analyze the derivative of $ \xi_{QQ} (r, z) $ with respect to redshift z, 
$ \frac{\partial \xi_{QQ}(r,z)}{\partial z}$, at any fixed $ r $, 
and find conditions which define its sign.
So long as the matter density correlation function in the range where 
\mbox{$0<x (r)\le \sigma^2(R) $}
(that corresponds for most models $r<30-40h^{-1}$Mpc) 
is monotonous function of $r$  its derivative with respect to redshift  z does not 
depend on value of $ r $. 
Therefore without losing of generality we can analyze this derivative analytically 
in the range \mbox{$ x (r) \ll1 $} where  
the expression (\ref{xi_qq}) becomes simpler: 
\begin{equation}
\label{xi_qq_t}
\xi_{QQ}^s(r,z)
=\nu^2(z)\cdot x(r).
\end{equation}
The  QSO correlation function (\ref{xi_pp}) in approach $ x(r) \ll 1$ will look as follows:
\begin{equation}
\label{xi_qq_tp}
\xi_{QQ}(r,z)
=\left(\nu(z)+\frac{\delta_c}{\nu(z)}\right)^2\cdot x(r).
\end{equation}
It means that  
$\frac{\partial \xi_{QQ}(r,z)}{\partial z} = 0$
when
\begin{equation}
\label{nu}
\nu(z_*)=\sqrt{\delta_c}.
\end{equation}
Taking into account (\ref{nu_z}) we shall obtain: 
\begin{equation}
\label{z*}
\sigma(R)={\sqrt{\delta_c}}\cdot(z_*+1).
\end{equation}

It means, that in the case when QSOs are formed in the peaks of the same scale $R$
at different $z$ their correlation function $\xi_{QQ}(r,z)$ monotonously increases
 $\left(\frac{\partial\xi_{QQ}(r,z)}{\partial z} > 0\right)$ if 
 $\nu(z) > \sqrt{\delta_c}$ and decreases $\left(\frac{\partial\xi_{QQ}(r,z)}{\partial z} < 0\right)$
if $\nu(z) < \sqrt{\delta_c}$. The interesting case is when condition (\ref{nu}) occurs
 at $z_*>0$. In this case the amplitude of $\xi_{QQ}(r,z)$ decreases in the range $0<z<z_*$, 
 reaches its minimum at $z=z_*$ and increases at $z>z_*$.   

In particular, in the case when the amplitude of $ \xi_{QQ}(r, z) $ decreases with increasing of $ z $,
or in the case of absence of essential evolution,
that was indicated by 
Iovino et al. 1991, Andreani, Cristiani 1992, Mo, Fang 1993, Komberg, 
Kravtsov 1995, it may mean that 
QSOs are formed in low peaks of a random Gaussian field of matter density
fluctuations, \mbox{$ \nu (z) \le\sqrt{\delta_c} $}. 
All these cases are shown in Fig.1. The solid line divides the plane into the ranges with decreasing
(left) and increasing (right) the amplitude of QSOs when $z$ increases. The moments at which the 
evolution changes the sign for standard CDM model, tilted ones and standard H+CDM 
(30\%Hot+60\%CDM+10\%baryon) one are shown by filled circles. 

\begin{figure}[t]
\epsfxsize=8.0cm
\epsfysize=30.0cm
\centerline{\epsfbox{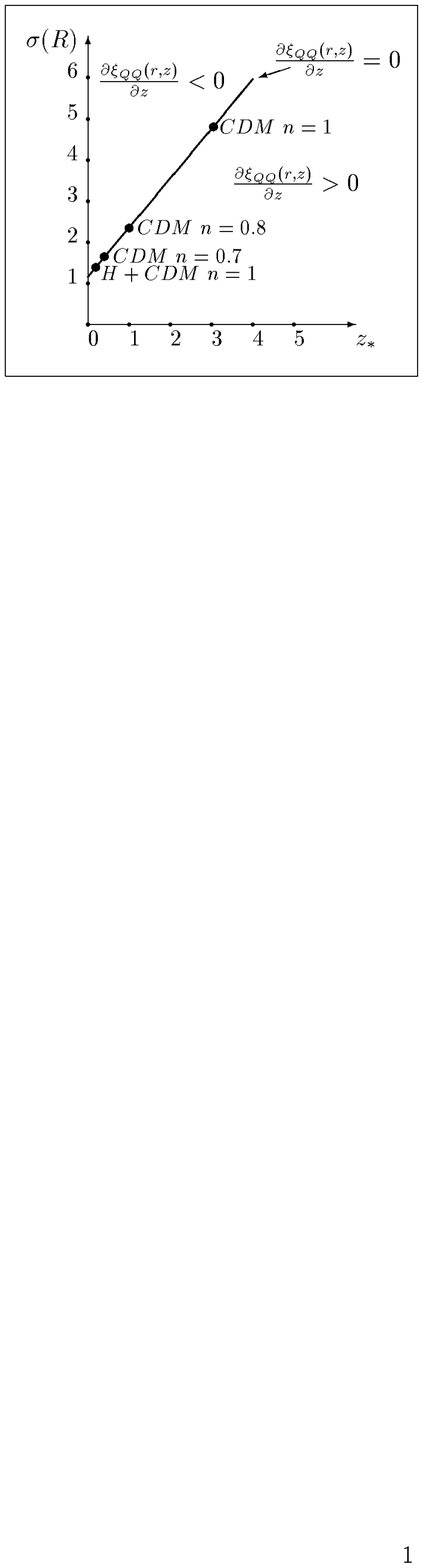}}
\vspace{-23cm}
\caption{
Izoline of minima of amplitudes
of QSO correlation function $\xi_{QQ}(r,z)$ in the plane $\sigma(R)-z$.
By filled circles it is shown the moments at which  the 
evolution of QSO correlation function changes the sign for 
for standard CDM model, tilted ones and standard H+CDM 
(30\%Hot+60\%CDM+10\%baryon) for the case when quasars are formed in the peaks of galaxy 
scale \mbox{$ R\sim0.35h^{-1} $ Mpc}.}
\end{figure}

The values of  $ z _ * $ for CDM  models with slopes of
their power spectra  $ n = 1,\;0.8,\;0.7 $ are  3.16, 1.06, 0.45 accordingly. 
For H+CDM model  $ z_* = 0.21 $. 
Thus, in the interval $ 0 < z < 3.16 $ the amplitude of $ \xi _ {QQ} (r, z) $ monotonously 
decreases only in CDM  model with $ n = 1 $ that agree with the data by 
Iovino et al. 1991; Andreani, Cristiani 1992; Mo, Fang 1993; Komberg, Kravtsov 1995. 
However, as  it is shown by Novosyadlyj, Chornij 1998 
owing to small height of peaks the amplitude of 
the correlation function of quasars $ \xi _ {QQ} (r, z) $ in this case becomes less 
than galactic one already at $ z\sim 1 $, that does not match mentioned above observational data.
It may mean that quasar phenomena would be connected with structures of larger scale than
galaxy one.

Therefore, let's analyze the case when scale of objects in which QSOs appear is different
at different redshifts and determine them using the last   data on the evolution of QSOs
correlation function $\bar \xi_{QQ}(r, z) $ given by La Franca et al. 1998 and 
Cristiani et al. 1998. It can be realised in the following way: a) we set some values 
for cosmological parameters that give us possibility to calculate the power spectrum
of density perturbations using the analytical approximation of H+CDM power spectra
in four dimensional parameter space (Novosyadlyj et al. 1999), b) we determine the $R(z)$
 using the $\bar \xi_{QQ}^{obs}(15h^{-1}{\rm Mpc},z)$
given by (\ref{xi_QQ_obs}) as input observable data at different $z$. Thereby we can determine 
the typical scales of structures $ R (z) $, in which the quasars at different $ z $ 
were formed. 
The scale $ R (z) $ can specify conditions and mechanism of formation of  QSOs: when
\mbox {$ R\sim0.5-1h^{-1} $ Mpc} it  suggests that main part of QSOs appeared 
 in the nuclei of massive galaxies which are not dynamically connected with structures
 of larger scales, 
when \mbox{$ R\ge 1\div2h^{-1}$ Mpc} the formation of quasars
in groups of galaxies due to processes of merges of galaxies is more probable.

The calculations were carried out  
for redshifts in the range $ 0.5\le z\le 3.5 $ and 
matter dominated flat  Universe with following values of the cosmological parameters:
$ \Omega_m = 1 $ (matter density in units of critical one), content of baryons $ \Omega_ b = 0.096$
followed from nucleosynthesis constraint (Tytler et al. 1996), 
content of the neutrino hot dark matter $ \Omega_{\nu} = 0.1 $ (one sort of 
massive neutrino), content of the cold dark matter  
$ \Omega _ {CDM} = \Omega_m-\Omega_b-\Omega_{\nu}$, dimensionless Hubble constant
$ h \equiv H_0/100(km/sec/Mpc)=0.5 $,
and different slopes of the primordial power spectrum - $ n = 0.8,1,1.2 $. 
The initial power spectra normalized to 4-year COBE data (Bennet et al. 1996, Bunn \& White 1997) 
are shown in Fig. 2. 
For each fixed value of redshift z we calculated the integrated correlation function of QSOs (2) using
the formulas (3, 4, 7), equated it to corresponding observational one (1) for $r=15h^{-1}$Mpc and solved
the obtained equation relative to $R(z)$. The results for three models are shown in Fig. 3.

\begin{figure}[th]
\epsfxsize=8.0cm
\centerline{\epsfbox{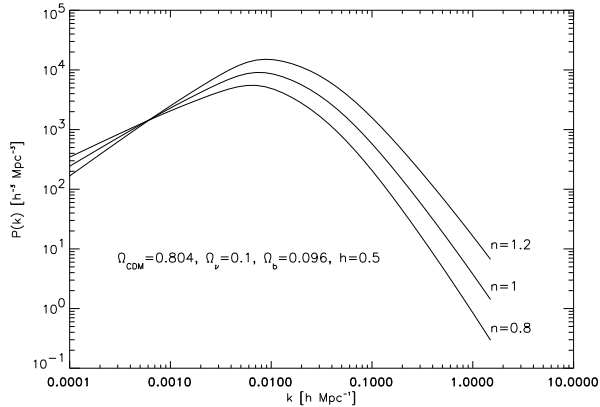}}
\vspace{0.0cm}
\caption{
Power spectra of density fluctuations   $ P (k) $ at $z=0$ (4-year COBE normalization) 
for cosmological
models with parameters $ \Omega_m = 1 $, $ \Omega_b = 0.096 $, $ \Omega_{\nu} = 0.1 $, 
$ \Omega_{CDM} = 0.804 $, $ h = 0.5 $, 
and different slopes of primordial power spectrum $ n = 0.8,1,1.2 $.}
\end{figure}

\begin{figure}[h]
\epsfxsize=8.0cm
\epsfysize=6.0cm
\centerline{\epsfbox{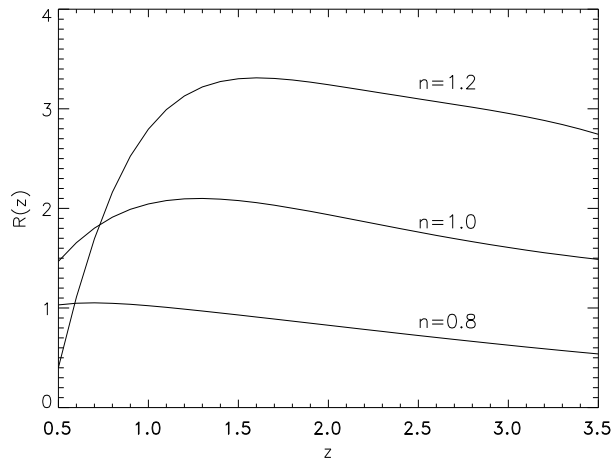}}
\vspace{0.0cm}
\caption{
The scales of structures $ R (z) $, in which were formed quasars at different 
$ z $, for cosmological models with the same parameters as in a fig.2.
}
\end{figure}

As one can see, in the models with low amplitude of the power spectrum at small scales ($ n = 0.8 $) 
the QSOs are formed in the peaks of galaxy scale  $ R = 0.5\div1$Mpc ($M\sim 10^{11}M_{\odot}$). 
And contrary, if the amplitude of the power spectrum is higher at small scales ($ n =1-1.2 $) then
the preferable range of scale of structures in which QSOs appeared  are 
$ R = 2\div3$Mpc, that corresponds to galaxy groups and clusters ($M\sim 10^{12}\div10^{13}M_{\odot}$).
It can be caused by merging of galaxies in their most dense regions at the center.
We do not discuss here the drastic decreasing of such scales at small redshifts for the model with $n=1.2$ 
because the analysis presented here has rather the qualitative character than quantitative. 
For the more certain quantitative predictions of scales of structures connected with QSOs phenomena 
the more realistic physical models of their formation must be taken into account.  
\\[2mm]

{\bf 4. Conclusions}\\[1mm]

 Based on the results presented here we conclude:

\begin{itemize}

\item The growth of the amplitude of the two-point  space correlation function of QSOs 
($\xi_{QQ}(r, z) $) with increasing of $ z $, which was 
revealed by La Franca et al. 1998; Cristiani et al. 1998 can be explained 
within the framework of the theory of formation of the large-scale structure 
of the Universe from the peaks of the random Gaussian  field  of matter density 
fluctuations.
The matter of explanation is that growth of statistical component of
$\xi_{QQ}(r, z)$  should be faster than the reduction of dynamical one.
 It can be provided by  choosing  of corresponding scale of structures $ R (z) $ in
which the QSOs were formed at different $ z $ .

\item The QSOs can be formed as in nuclei of single massive galaxies as in 
consequence of merging of galaxies in their groups or clusters. Which mechanism
is statistically dominant it depends on an amplitude of the initial power spectrum 
at small scale.  
\\[2mm]
\end{itemize}

{\bf References\\[1mm]}
Andreani P., Cristiani S.: 1992, {\it Ap. J.},
{\bf 398}, L.13.\\
Bardeen J.M., Bond J.R., Kaiser N., Szalay A.S.: 1986,
{\it Ap. J.}, {\bf 304}, 15.\\
Bennett, C.L., Banday, A.J., Gorski, K.M., \h  Hinshaw, G., Jackson, P., Keegstra, P., Kogut, A.,\h  Smoot, G.F., Wilkinson, D.T., Wright, E.L.: 1996,\h {\it ApJL}, {\bf 464}, L1.\\
Bunn, E.F., \& White, M.: 1997, {\it ApJ}, {\bf 480}, 6.\\
Cristiani S., La Franca F., Andreani P.: {\it astro-ph/9802291}.\\
Efstathiou G., Rees M.: 1988, {\it MNRAS}, {\bf 230}, 5.\\
Fang Li-Zhi: 1998, {\it astro-ph/9804106}.\\
Haehnelt M.: 1993, {\it MNRAS}, {\bf 265}, 727.\\
Haehnelt M., Rees M.J.: 1993, {\it MNRAS}, 
{\bf 263}, 168.\\
Iovino A., Shaver P., Cristiani S.: 1991, 
{\it in The Spase Distribution of Quasars, 
ed. D. Crampton (ASP Conf.Ser.)}, {\bf21}, 202.\\
Kaiser N.: 1984, {\it Ap. J.}, {\bf 284}, L.9.\\
Kontorovich V.M.: 1996, {\it Astr. Astrophys. Trans.}, 
{\bf 10}, 315.\\
Komberg B.V., Kravtsov A.V.: 1995, 
{\it Astr. Astrophys. Trans.}, {\bf 8}, 241.\\
La Franca F., Andreani  P., Cristiani S.:
{\it astro-ph/9802281}.\\
Loeb A., Rasio F.A.: 1994, {\it Ap. J.}, {\bf 432}, 52.\\
Mo H.J., Fang L.Z.: 1993, {\it Ap. J.}, {\bf 410}, 493.\\
Novosyadlyj B., Chornij Yu.: 1998, {\it JPS}, {\bf 2}, 433.\\
Novosyadlyj B., Chornij Yu.: 1998, {\it KFNT}, {\bf 14}, 156.\\
Novosyadlyj B.: 1994, {\it KFNT}, {\bf 10}, 13.\\
Novosyadlyj B.S.: 1996, {\it Astr.Astrophys.Trans.},
{\bf 10}, 85.
Novosyadlyj B.S., Durrer R. and Lukash V.N.: 1999, A\&A, 347, 799.\\
Nusser A., Silk J.: 1993, {\it Ap. J.}, {\bf 411}, L1.\\
Tytler D., Fan X.M., \& Burles S. 1996, Nature, 381, 207.\\
\\[3mm]\indent
\end{document}